\documentclass[twocolumn,showpacs,amsmath,amssymb,aps,prd,floats,nofootinbib]{revtex4-2}
\usepackage{graphicx}
\usepackage[switch,pagewise,running]{lineno}
\usepackage{dcolumn}
\usepackage{bm}
\usepackage{epsfig}
\usepackage{xspace}
\usepackage{hyperref}
\hypersetup{colorlinks,linkcolor=blue,citecolor=blue,urlcolor=blue}  
\usepackage{color,xcolor}
\usepackage{verbatim}

\begin{document}
\title{A unified model for orphan and multi-wavelength blazar flares}
\author{Ze-Rui Wang$^{1,2}$, Ruo-Yu Liu$^{1,2}$, Maria Petropoulou$^3$, Foteini Oikonomou$^4$, Rui Xue$^5$, and Xiang-Yu Wang$^{1,2}$}
\affiliation{$^1$School of Astronomy and Space Science, Nanjing University, 210023 Nanjing, Jiangsu, China; \textcolor{blue}{ryliu@nju.edu.cn}\\
$^2$Key Laboratory of Modern Astronomy and Astrophysics (Nanjing University), Ministry of Education, Nanjing 210023, China\\
$^3$Department of Physics\\
National and Kapodistrian University of Athens, GR-157 84 Zografou, Greece; \textcolor{blue}{mpetropo@phys.uoa.gr}\\
$^4$Institutt for fysikk, NTNU, Trondheim, Norway; \textcolor{blue}{foteini.oikonomou@ntnu.no}\\
$^5$Department of Physics, Zhejiang Normal University, Jinhua 321004, China}

\begin{abstract}
Blazars are a class of active galactic nuclei which host relativistic jets oriented close to the observer's line of sight. Blazars have very complex variability properties. Flares, namely flux variations around the mean value with a well-defined shape and duration, are one of the identifying properties of the blazar phenomenon. Blazars are known to exhibit multi-wavelength flares, but also ``orphan" flares, namely flux changes that appear only in a specific energy range. Various models, sometimes at odds with each other, have been proposed to explain specific flares even for a single source, and cannot be synthesized into a coherent picture. In this paper, we propose a unified model for explaining orphan and multi-wavelength flares from blazars in a common framework. We assume that the blazar emission during a flare consists of two components: (i) a quasi-stable component that arises from the superposition of numerous but comparatively weak dissipation zones along the jet, forming the background (low-state)  emission of the blazar, and (ii) a transient  component, which is responsible for the sudden enhancement of the blazar flux, forming at a random distance along the jet by a strong energy dissipation event. Whether a multi-wavelength or orphan flare is emitted depends on the distance from the base of the jet where the dissipation occurs. Generally speaking, if the dissipation occurs at a small/large distance from the supermassive black hole, the inverse Compton/synchrotron radiation dominates and an orphan $\gamma$-ray/optical flare tends to appear. On the other hand, we may expect a multi-wavelength flare if the dissipation occurs at an intermediate distance. We show that the model can successfully describe the spectral energy distribution of different flares from the flat spectrum radio quasar 3C~279 and the BL~Lac object PKS~2155-304.
\end{abstract}

\maketitle

\section{Introduction}
Blazars are the most extreme subclass of active galactic nuclei (AGN) with a relativistic jet closely aligned to the observer's line of sight  \citep{1984RvMP...56..255B, 1995PASP..107..803U}. Blazars are divided into flat-spectrum radio quasars (FSRQs) and BL~Lacertae objects (BL~Lacs) based on the presence or not of strong broad emission lines in their optical-ultraviolet spectra \citep{1995PASP..107..803U}. The spectral energy distribution (SED) and multi-wavelength flux variability of blazars are important tools for studying the physics of extragalactic relativistic jets.

The SED of a blazar is dominated by the non-thermal radiation of the jet, and exhibits a characteristic double-hump structure \citep{1998MNRAS.299..433F}. The low energy component is believed to arise from the synchrotron radiation of relativistic electrons in the magnetic field, while the high-energy component is probably produced by inverse Compton (IC) scattering of relativistic electrons off low-energy target photons. These could be synchrotron photons produced by the same electron population in the jet (SSC model, \cite[e.g.][]{1996ApJ...461..657B,1996A&AS..120C.537M}), or could be external photons (EC model, \cite[e.g.][]{1992A&A...256L..27D,1993ApJ...416..458D,1994ApJ...421..153S}).

A significant enhancement of a blazar's flux, usually considered to be a factor of $2-3$~above its average value, is referred to as a blazar flare. During flares the polarization and the spectral index of a blazar's emission may also experience dramatic variability \cite[e.g.][]{2020MNRAS.492.1295P}. Blazar flares have been commonly observed at different energy bands (e.g., from radio wavelengths to very high-energy $\gamma$-rays) with different duration timescales ranging from several years to a few minutes \cite[e.g.][]{1990ApJ...356..432G, 2007AJ....133.1947N,2008A&A...486..411S,2008ApJ...677..906F,2008MNRAS.384L..19B,2009ApJ...691L..13D,2010ApJ...722..520A,2011ApJ...738...25A,2012ApJ...756...13B,2015ApJ...807...79H}. Based on their duration, one can roughly define three types of blazar flares, namely year-long flares \cite[e.g.][]{2013A&A...553A.107C}, day-to-month long flares \cite[e.g.][]{2011ApJS..192...12I} and intraday flares \cite[e.g.][]{1995ARA&A..33..163W}. Different duration timescales may relate to different aspects of blazar physics. For instance, year-long flares are thought to be related to the orbital motion of a binary or jet precession \citep{2004A&A...419..913O,2018MNRAS.478.3199B}, or instabilities in the accretion flow \citep{2010MNRAS.402.2087V, 2011MNRAS.418L..79T}. Day-to-month-long and intraday flares are thought to be related to processes taking place in the jet itself, and can be reasonably explained by models involving one or more compact dissipation zones in the jet \cite[e.g.][]{2012AJ....143...23G, 2013A&A...557A..71R}. Therefore, measurements of day-to-month-long and intraday flares, including their temporal variability and spectrum, could constrain the physical parameters of the dissipation zone, such as size, location, geometry, bulk Lorentz factor, particle acceleration, and cooling processes.

One of the most peculiar aspects of blazar variability are the so-called orphan flares. These are flares that occur in a specific energy band without correlated variability in other bands, and have been discovered in many blazars (orphan X-ray flare \cite[e.g.][]{2013A&A...552A..11R}; orphan optical flare \cite[e.g.][]{2013ApJ...763L..11C,2019hepr.confE..27W,2019Galax...7...21W, 2019ApJ...880...32L}; orphan GeV flare \cite[e.g.][]{2019ApJ...880...32L, 2019ApJ...884..116L,2017ApJ...836..205A,2016MNRAS.463L..26B,2015ApJ...804..111M,2017ApJ...850...87M}; orphan TeV flare \cite[e.g.][]{2017ApJ...836..205A}). Interestingly, different types of flares, i.e., orphan flares and multi-wavelength flares, have been observed to occur in the same blazar from time to time. For instance, the FSRQ PKS~0208-512 exhibited three flares at optical and near-infrared wavelengths within 3 years, with the second one having no $\gamma$-ray counterpart. Ref.~\citep{2013ApJ...771L..25C} found that the Compton dominance ($q$), which is defined as the luminosity ratio between the IC component and synchrotron component, was different for the three flares. This was interpreted as evidence for a varying magnetic field and/or varying soft photon field during these optical outbursts.

Various models have been proposed to explain the origins of orphan flares. For example, Ref.~\citep{2016MNRAS.463L..26B} proposed that the orphan $\gamma$-ray flare of PKS~1222+21 can be explained when relativistic blobs in the jet encounter luminous stars. Ref.~\citep{2015ApJ...804..111M} suggested that the ring-of-fire model (a blob propagates relativistically along the fast-moving spine of a blazar jet and passes through a synchrotron-emitting ring of electrons from the slow-moving sheath of the jet) can reproduce the orphan $\gamma$-ray flare of PKS~1510-089. Ref.~\citep{2014IJMPS..2860180C} suggested the orphan optical flare of PKS~0208-512 can be explained by different allocations of energy between the magnetization of the emitting region and particle acceleration. Ref.~\citep{2016Galax...4...45J} showed that the orientation of the magnetic field might be associated with orphan flares. Ref.~\citep{2021MNRAS.503..688S} argues that an orphan $\gamma$-ray flare from FSRQs is likely to arise if the particles are accelerated in magnetically dominated pair plasmas. The synchrotron emission is suppressed since the particles are accelerated nearly along the direction of the local magnetic field (small pitch angles), while the $\gamma$-ray flare is produced by inverse Compton scattering on an external radiation field. Ref.~\citep{2005ApJ...630..186R} explained an orphan TeV flare of 1ES 1959+650 with a hadronic model in which relativistic protons interact with the photon field supplied by electron synchrotron radiation reflected off a dilute reflector. While all these models are viable and can reproduce the spectral features of an orphan flare, one may wonder if there is a single scenario that can apply to all orphan flares. More importantly, if both orphan and multi-wavelength flares are observed from the same blazar, do we have to apply different models to explain different types of flares or can a single scenario account for both? 

In this work, we attempt to establish a connection between orphan and multi-wavelength flares occurring in a certain blazar, and search for a theoretical interpretation of the spectral variety of blazar flares in a unified physical picture. In general, the non-thermal blazar emission is produced when the jet's energy (magnetic or kinetic) is dissipated and transferred to relativistic particles. The properties of the resulting non-thermal emission may strongly depend on the distance of the dissipation site from the central supermassive black hole (SMBH), since the physical environment can experience a pronounced change along the jet \cite[e.g.][]{1979ApJ...232...34B, 1980ApJ...235..386M, 1989ApJ...340..181G, 2013ApJ...771L..25C}. However, the location of the dissipation zone in blazar jets remains uncertain \cite[e.g.][]{2011A&A...534A..86T,2016ARA&A..54..725M, 2019BAAS...51c..92R}. In some previous studies, it was suggested that dissipation may occur stochastically along the jet of a blazar \cite[e.g.][]{2019PhRvD..99f3008L, 2019ApJ...886...23X, 2020arXiv201210291P}. In this framework, the non-flaring emission of a blazar results from the superposition of radiation produced in numerous dissipation zones. If additional energy dissipation takes place in one (or a few) of them, so that its (their) emission outshines the rest of the jet, then the blazar is expected to flare. Therefore, there may be no essential difference between the non-flaring state and the flaring state of a blazar, except that the flaring state is related to a much stronger dissipation event. The distance of the flaring zone to the SMBH can then determine the spectral and temporal properties of the flare. In this paper, we explore in detail such a scenario. Hereafter, we refer to it as the stochastic dissipation model.

The rest of the paper is organized as follows. The model setup is introduced in Section~\ref{sec:method}. In Section~\ref{sec:sec3} we apply the model to two well-known blazars, namely the FSRQ 3C~279 and the BL~Lac PKS~2155-304. In Section~\ref{sec:sec4}, we study the applicability of the model to the general blazar population. The discussion and conclusions are given in Section~\ref{sec:sec5} and Section~\ref{sec:sec6}, respectively. In this work, we use the following cosmological parameters, $H_{0}= 68$ km $\rm s^{-1} Mpc^{-1}$, $\Omega_{\Lambda} = 0.7$ and $\Omega_{\rm M} = 0.3$. 

\section{Model Setup} \label{sec:method}
We assume that the emission of the blazar jet in a flaring state is composed of at least two emission components. One component arises from a flaring zone that dominates the flare emission. The appearance of such components may be related to MHD instabilities \citep{kink, current_driven} or magnetic reconnections in the jet  \citep{1998ApJ...493..291B,2009MNRAS.394L.126M,2009MNRAS.395L..29G,2013MNRAS.431..355G,2016MNRAS.462.3325P}. The other component is the superposition of emission from numerous but comparatively weak dissipation zones along the entire jet. The latter can be regarded as a background emission to the flare \cite[e.g.][]{2019ApJ...877...39M} and might describe the non-flaring blazar emission. The two emission components are assumed to be decoupled from each other. Since we focus primarily on the former component in this work, the modelling of the background radiation spectrum is simply characterized with a polynomial function. 

\begin{figure*}[htbp]
\centering
\includegraphics[width=2\columnwidth]{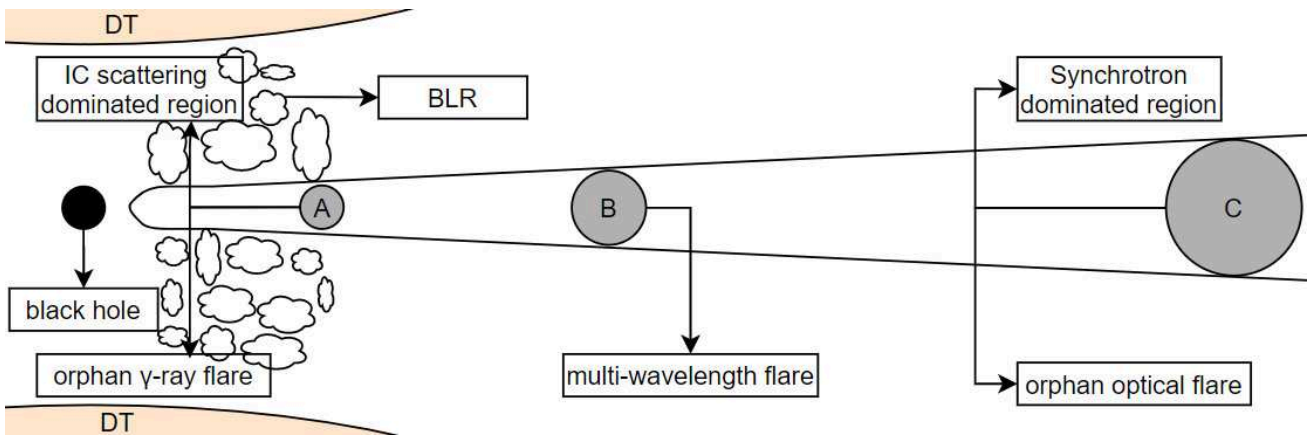}
\caption{Schematic view (not to scale) of the stochastic dissipation model. DT and BLR represent dusty torus and broad-line region respectively. The background radiation comes from numerous but relatively weak dissipation zones (not shown here). The flaring zone (indexed blobs) occurs at a random distance from the base of the jet. We argue that the orphan $\gamma$-ray flares are more likely to arise, if the flaring zone occurs in location A. The orphan optical flares may arise if the flaring zone occurs in location C, while multi-wavelength correlated flares are expected in location B.  \label{fig:sketch}}
\end{figure*}
In Fig.~\ref{fig:sketch} we show a sketch of the considered scenario. Since the ratio between the power of the synchrotron radiation and the power of the IC radiation is roughly proportional to the ratio between the energy density of the magnetic field $u_{\rm B}$ and that of the target radiation field $u_{\rm ph}$, the emission from the flaring zone will be dominated by the IC process if it occurs relatively close to the SMBH, given the presence of the broad line region (BLR) and/or the dusty torus (DT). As a result, the $\gamma$-ray emission from the flaring zone may exceed that of the jet's background emission, while the synchrotron radiation of the flaring zone at lower frequency could still be subdominant. In this case, the blazar's emission is enhanced specifically at the $\gamma$-ray band, and the blazar appears to be experiencing an orphan $\gamma$-ray flare. As the distance of the flaring zone to the SMBH increases, the synchrotron radiation becomes increasingly important with respect to either SSC or EC radiation. The synchrotron process could dominate if the dissipation takes place beyond a certain distance and an orphan optical flare may then be expected.

For a quantitative description, we assume that electrons are continuously injected into the flaring zone (as long as this remains active) with a broken power-law energy distribution \cite[e.g.][]{2009MNRAS.397..985G, 2010MNRAS.402..497G}
\begin{equation}\label{eq:1}
n_{\rm e}^{\rm inj}(\gamma)\propto\left\{
\begin{array}{ll}
\gamma^{-p_1}, & \gamma_{\rm min}\leq \gamma\leq  \gamma_{\rm break}~\\
\gamma_{\rm break}^{p_2-p_1}\gamma^{-p_2}, & \gamma_{\rm break}<\gamma\leq\gamma_{\rm max},
\end{array}
\right.
\end{equation}
where $\gamma_{\min}, \gamma_{\max}$ are the minimum and maximum Lorentz factors of the distribution, $\gamma_{\rm break}$ is the break Lorentz factor, which is not related to radiative cooling, $p_1$ and $p_2$ are respectively the low-energy slope and the high-energy slope of the broken power-law spectrum. Assuming a spherical dissipation zone with radius $R$ in the comoving frame and given the electron injection luminosity $L_{\rm e}^{\rm inj}$, the steady-state electron density distribution can then be written as \citep{2019ApJ...886...23X}
\begin{equation}
N_{\rm e}(\gamma)=\frac{3L_{\rm e}^{\rm inj}n_{\rm e}^{\rm inj}(\gamma)}{4\pi R^3m_{\rm e}c^2 \int{\gamma n_{\rm e}^{\rm inj}(\gamma){\rm d}{\gamma}}}{\rm min}(t_{\rm cool}(\gamma),t_{\rm esc}),
\end{equation}
where $t_{\rm esc}=R/c$ is the electron escape timescale, $t_{\rm cool}=3m_{\rm e}c/(4\sigma_{\rm T}\gamma(u_{\rm B}+f_{\rm KN}u_{\rm ph}))$\footnote{$u_{\rm ph}$ here does not count in the density of the synchrotron radiation. Taking it into account makes the calculation become non-linear \citep{2010A&A...524A..31Z}. Although it may be dealt with iteratively, it will lead to an excessively expensive computation when we apply the MCMC method to the spectral fitting later. To evaluate the influence, we have re-calculated the model flux for the best-fit parameters shown in Table~\ref{tab:parameters}, after including the synchrotron radiation in $u_{\rm ph}$ when calculating the cooling of electrons. The difference is 10\% at most (for flare 1 of PKS~2155-304 without external radiation field) for $E_\gamma<1\,$TeV.} is the electron radiative cooling timescale, $\sigma_{\rm T}$ is the Thomson scattering cross section, and $f_{\rm KN}$ is the factor accounting for Klein-Nishina (KN) effects \citep{2005MNRAS.363..954M}. The synchrotron and IC emission from the relativistic electrons can be then calculated given the magnetic field and the radiation field. High-energy photons from the IC process can be absorbed by soft photons via photon-photon pair production. We consider the absorption of $\gamma$-ray photons due to the synchrotron radiation of the flaring zone, the BLR radiation and the DT radiation, as well as the extragalactic background light (EBL, model C in Ref.~\citep{2010ApJ...712..238F}) during the propagation in the intergalactic space.

The magnetic field strength in the dissipation zone may vary with the distance to the SMBH, $r$. Considering a truncated conical jet and assuming that the radius of the dissipation zone $R$ is comparable to the transverse radius of the jet at its location, we may write
\begin{equation}\label{eq:2}
R(r) = R_0\left(\frac{r}{0.1~\rm pc}\right),
\end{equation}
where $R_0$ is the transverse radius of the jet at 0.1 pc. If we assume a constant magnetic luminosity along the jet, which is consistent with some results of the VLBA survey \cite[e.g.][]{2009MNRAS.400...26O,2011A&A...532A..38S}, the magnetic field strength can also be parameterized as a function of $r$
\begin{equation}\label{eq:3}
B(r)=B_0\frac{R_0}{R(r)}, 
\end{equation}
where $B_0$ is the magnetic field strength of the dissipation zone for $r=0.1$\,pc.

The target radiation field for the IC process consists of the synchrotron radiation of the electrons in the dissipation zone and the external photon field. Both radiation fields depend on the distance $r$ of the flaring zone to the SMBH. The energy density of synchrotron photons (as measured in the comoving frame of the dissipation zone) is given by $u_{\rm syn}(r) = L_{\rm syn}/(4\pi c R(r)^2\Gamma^4)$, where $L_{\rm syn}$ is the luminosity of synchrotron emission in the observer's frame and $\Gamma$ is the bulk Lorentz factor of the jet. The energy density of external photons, i.e., from the BLR and DT, in the jet comoving frame can be written as
\cite{2012ApJ...754..114H}
\begin{equation}\label{eq:5}
u_{\rm BLR}(r)=\frac{\xi_{\rm BLR}\Gamma^2 L_{\rm disk}}{3\pi r^2_{\rm BLR}c[1+(r/r_{\rm BLR})^{\beta_{\rm BLR}}]}
\end{equation}
\begin{equation}\label{eq:6}
u_{\rm DT}(r)=\frac{\xi_{\rm DT}\Gamma^2 L_{\rm disk}}{3\pi r^2_{\rm DT}c[1+(r/r_{\rm DT})^{\beta_{\rm DT}}]},
\end{equation}
where $\xi_{\rm BLR}=0.1$ and $\xi_{\rm DT}=0.1$ are the fractions of the disk luminosity reprocessed into BLR and torus radiation, respectively, $r_{\rm BLR}=0.1(L_{\rm disk}/10^{46}\rm erg~s^{-1})^{1/2}$ pc and $r_{\rm DT}=2.5(L_{\rm disk}/10^{46}\rm erg~s^{-1})^{1/2}$ pc denote the characteristic distances of the BLR and torus from the SMBH in the AGN frame. We assume that the radiation energy density drops steeply with distance beyond the characteristic distance $r_{\rm BLR(DT)}$, adopting $\beta_{\rm BLR}=3$ \cite{2009ApJ...704...38S} and $\beta_{\rm DT}=4$ \cite{2012ApJ...754..114H}. The spectral shape of the radiation fields is assumed to be that of a grey body peaking at a frequency $4.5\times10^{14}\Gamma$ Hz for the BLR \cite{2019PhRvD..99f3008L} and at $3\times 10^{13}\Gamma$ Hz for the DT \cite{2009MNRAS.397..985G}, both measured in the jet's comoving frame. Note that the background emission of the jet (or the non-flaring emission) may also serve as a target photon field for the IC process. Its influence is, however, minor as will be shown in Section~\ref{sec:sec5}. For simplicity, we ignore it as a target photon field in the following calculations.  

There have been suggestions that the jet decelerates from highly relativistic speeds to mildly or sub-relativistic speeds on kiloparsec scales \citep{1997MNRAS.286..425W,1999MNRAS.304..135H,2008A&A...491..321M,2014MNRAS.441.1488P}. Continuous jet models involving decelerating flows have been used to fit the SEDs of AGN \citep{2003ApJ...594L..27G,2009ApJ...699...31A,2013MNRAS.429.1189P}. Following Ref.~\cite{2013MNRAS.429.1189P}, we assume the jet's bulk Lorentz factor remains constant up to 0.1\,pc as $\Gamma_0\gg 1$, and decelerates beyond this distance as a function of $\log(r)$, reaching $\Gamma_{\rm min}=2$ at 100\,pc. We approximate the Doppler factor by $\delta_{\rm D}\approx \Gamma$. Then the Doppler factor for $r>0.1~{\rm pc}$ can be given by \citep{2013MNRAS.429.1189P}:
\begin{equation}\label{eq:4}
\delta_{\rm D}(r)=\delta_{\rm D,0} - \frac{\delta_{\rm D,0}-2}{{\rm log}(\frac{100~{\rm pc}}{0.1~{\rm pc}})}{\rm log}\left(\frac{r}{0.1~\rm pc}\right),
\end{equation}
where $\delta_{\rm D,0}\approx \Gamma_{0}$ is the Doppler factor at 0.1 pc, noting that other forms of $\delta_D(r)$ may also be possible. 
In fact,
some observations  \cite[e.g.][]{2009ApJ...706.1253H,2012ApJ...758...84P,2013AJ....145...12B,2015ApJ...798..134H, 2019ApJ...887..147P}, numerical simulations \cite[e.g.][]{2007MNRAS.380...51K} and theoretical studies \citep[e.g.][]{2019MNRAS.484.1378G} show that the jet may still accelerate from sub-parsec up to tens of parsec scales. We will discuss a constant-speed jet case and an accelerating jet case in Sec.~\ref{sec:sec4}. 

In total, there are ten free parameters for one dissipation zone. Among them, six parameters are related to the spectrum of electron injection, i.e., three for the characteristic electron Lorentz factors ($\gamma_{\rm min}$, $\gamma_{\rm break}$, and $\gamma_{\rm max}$), two for the power-law slopes of the electron spectrum ($p_1$ and $p_2$), and one for the electron injection luminosity ($L_{\rm e}^{\rm inj}$). 
The remaining four parameters are related to the physical properties of the flaring zone, namely the distance of the flaring zone from the SMBH $r$, the radius of the flaring zone $R$, the magnetic field strength $B$, and the Doppler factor $\delta_{\rm D}$. For the modeling of multiple flares from the same source, the last three parameters are not independent of each other, but are related to their distance from the black hole $r$ and their values at 0.1 pc ($R_0$, $B_0$, $\delta_{\rm D,0}$). To reduce the number of free parameters, we fix the values of $p_1$, $\gamma_{\rm min}$ and $\gamma_{\rm max}$ in different flares of a given blazar.

To summarize, for the modeling of multiple flares from one source, our model has six parameters ($R_0$, $B_0$, $p_1$, $\gamma_{\rm min}$, $\gamma_{\rm max}$, $\delta_{\rm D,0}$) that are common among different dissipation sites, and four parameters ($L_{\rm e}^{\rm inj}$, $p_2$, $\gamma_{\rm break}$, $r$) that are unique to each dissipation site. Thus, if we apply the stochastic dissipation model to explain, for instance, three flares from a blazar, we have to specify in total eighteen parameters.

\section{Application to 3C~279 and PKS~2155-304} \label{sec:sec3}
Blazars are historically divided into two classes, namely FSRQs and BL~Lacs, according to their optical spectra. The former display strong, broad emission lines, while the latter show at most weak emission lines, and in many cases are completely featureless \citep{1995PASP..107..803U}. These sources are thought to be powered by accretion disks with different mass accretion rates and radiative efficiencies (for a review, see Ref.~\citep{2019ARA&A..57..467B}). As a result, the strength of ambient photon fields in these blazar subclasses is expected to be very different. 

\begin{table*}
\caption{\label{tab:parameters}Best-fit parameters for three flares of 3C~279 and PKS~2155-304 modeled with the stochastic dissipation scenario. Errors correspond to the $1\sigma$ uncertainties.}
\begin{ruledtabular}
\begin{tabular}{c|ccc|ccc}
                            & \multicolumn{3}{c|}{3C~279}                                                                                 & \multicolumn{3}{c}{PKS~2155-304}                                                        \\  \hline
$L_{\rm disk}$ ($10^{42}~\rm erg~s^{-1}$) & \multicolumn{3}{c|}{$2000$\footnote{The disk luminosity of 3C~279 is $2\times 10^{45}~\rm erg~s^{-1}$ \citep{1999ApJ...521..112P}.}}                                                                          &  \multicolumn{3}{c}{1\footnote{Ref.~\citep{2011ApJ...732..113S} has measured and found no emission line is observed in the spectrum of PKS~2155-304. According to their analysis, the BLR luminosity upper limit of PKS~2155-304 is $1.1\times10^{41}~\rm erg~s^{-1}$. So we set the BLR luminosity of $10^{41}~\rm erg~s^{-1}$ and calculate the corresponding disk luminosity of $10^{42}~\rm erg~s^{-1}$.}} \\
$R_0$
($10^{16}~$cm)                      & \multicolumn{3}{c|}{$0.91_{-0.08}^{+0.07}$\footnote{This corresponds to a half-opening angle of for $\sim1.7^{\circ}$ the truncated conical jet.
}}                                                                           & \multicolumn{3}{c}{$3.41_{-0.54}^{+0.88}$\footnote{This corresponds to a half-opening angle of for $\sim6.3^{\circ}$ the truncated conical jet.
}}                                            \\ 
$B_0$
(G)                       & \multicolumn{3}{c|}{$0.45_{-0.04}^{+0.03}$}                                                                                          & \multicolumn{3}{c}{$0.14_{-0.01}^{+0.02}$}                                                           \\ 
$\gamma_{\rm min}$ ($10^2$)                         & \multicolumn{3}{c|}{$5.07_{-0.06}^{+0.05}$}    & \multicolumn{3}{c}{$1.01_{-0.34}^{+0.68}$}                \\ 
$\gamma_{\rm max}$\footnote{The theoretical limit of $\gamma_{\rm max}$ can be obtained by equating the acceleration and the cooling or escape timescales, i.e., $t_{\rm acc}={\rm min}(t_{\rm cool},t_{\rm esc})$, where $t_{\rm acc}\sim E/ecB$ can be given in the limiting case following Ref.~\citep{2002PhRvD..66b3005A}. We obtain $\gamma_{\rm max}=9.9\times 10^{6} $ for 3C~279 and $\gamma_{\rm max}=7.9\times 10^{7} $ for PKS~2155-304 based on the best-fit parameters. The best-fit values of $\gamma_{\rm max}$ given by the MCMC method are consistent with these theoretical limits.} ($10^6$)          & \multicolumn{3}{c|}{$9.68_{-0.65}^{+45.65}$}                                                                             & \multicolumn{3}{c}{$1.56_{-0.47}^{+0.78}$}                                               \\ 
$p_{\rm 1}$                 & \multicolumn{3}{c|}{$1.80_{-0.04}^{+0.06}$}                                                                                          & \multicolumn{3}{c}{$1.63_{-0.04}^{+0.03}$}                                                           \\ 
$\delta_{\rm D,0}$ &
 \multicolumn{3}{c|}{$70.6_{-3.4}^{+4.5}$}                                                                                           & \multicolumn{3}{c}{$53.6_{-7.0}^{+6.7}$}                                                            \\ 
  \hline Flare state       & Flare 1            & Flare 2            & Flare 3                                          & Flare 1            & Flare 2            & Flare 3            \\
  \hline
$L_{\rm e}^{\rm inj} (10^{44}~\rm erg~s^{-1})$    & $80.0_{-8.1}^{+10}$ & $943_{-55}^{+53}$ & $314_{-52}^{+55}$ &  $0.32_{-0.05}^{+0.07}$ & $0.21_{-0.04}^{+0.08}$ & $8.74_{-2.27}^{+3.33}$ \\ 
$p_{\rm 2}$                                                                                             & $3.75_{-0.16}^{+0.14}$               & $3.37_{-0.05}^{+0.05}$                & $3.29_{-0.15}^{+0.22}$     & $3.72_{-0.15}^{+0.11}$                  & $3.24_{-0.10}^{+0.06}$                & $4.27_{-0.07}^{+0.06}$              \\ 
$\gamma_{\rm break} (10^4)$        &                                     $0.34_{-0.04}^{+0.03}$    & $0.43_{-0.07}^{+0.07}$    & $4.15_{-1.27}^{+1.57}$    & $2.50_{-0.57}^{+0.41}$    & $3.88_{-0.75}^{+1.73}$    & $6.55_{-0.52}^{+0.68}$     \\ 
$r$ ($10^{-1}~$pc)                                                                                      & $2.01_{-0.17}^{+0.23}$               & $30.8_{-1.9}^{+1.7}$                & $125_{-18}^{+22}$                      & $0.12_{-0.01}^{+0.01}$                & $0.51_{-0.07}^{+0.10}$                & $39.2_{-3.3}^{+2.3}$            
\end{tabular}
\end{ruledtabular}
\end{table*}

\begin{figure*}[htbp]
\centering
\includegraphics[width=2\columnwidth]{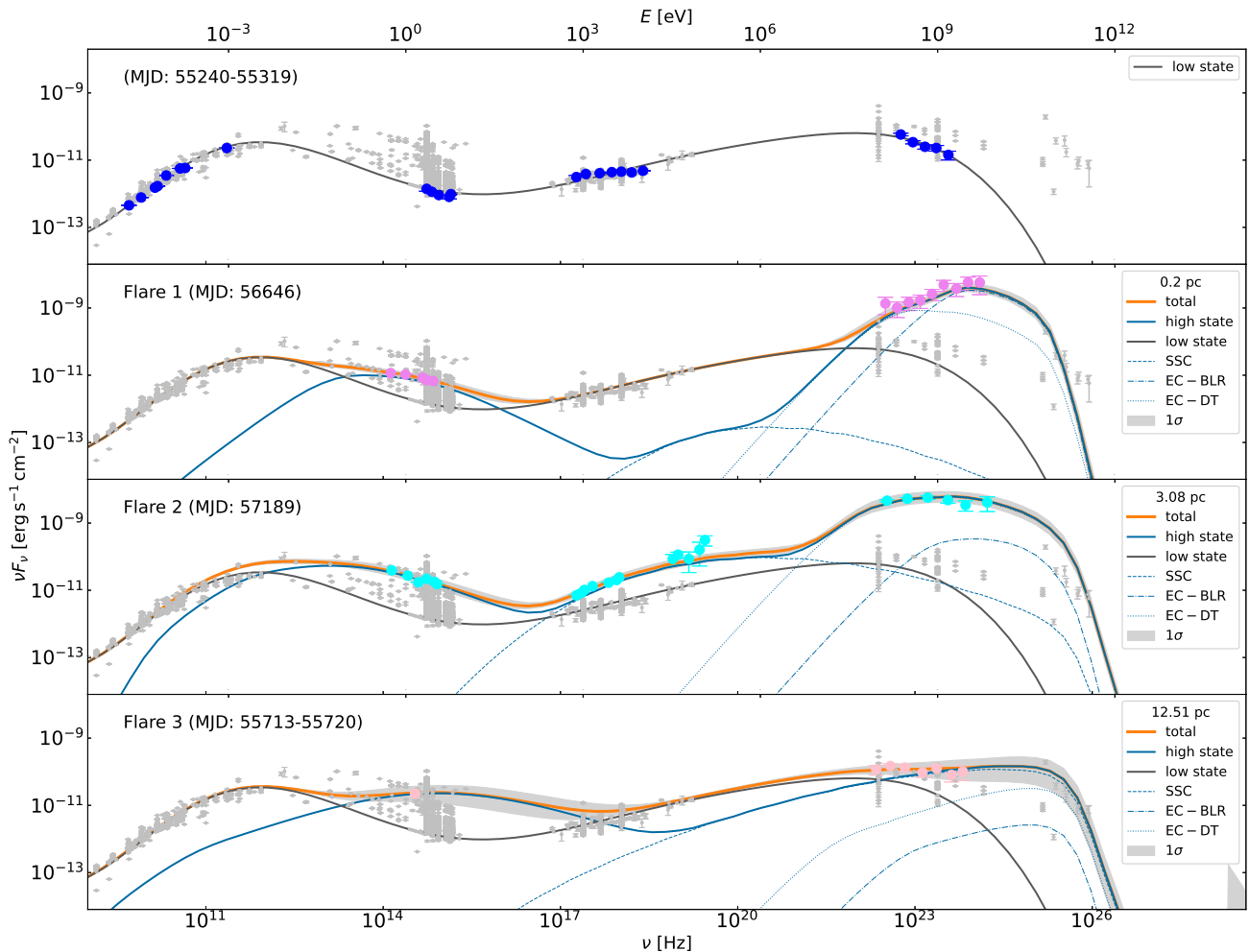}
\caption{The fitting results for 3C~279. The black solid lines represent background radiation from many dissipation zones, and the orange solid lines are total radiation including background radiation and the emission from a flaring zone. The blue solid lines represent the total emission from the flaring zone. The dashed, dot-dashed and dotted lines are IC emission from the flaring zone for different seed photon fields (see inset legends). The grey points show archival data, and the colored symbols show the data points that correspond to the three different states of 3C~279. The references for the data can be found in Section~\ref{sec:sec3}.
\label{fig:3C 279}}
\end{figure*}

\begin{figure*}[htbp]
\centering
\includegraphics[width=2\columnwidth]{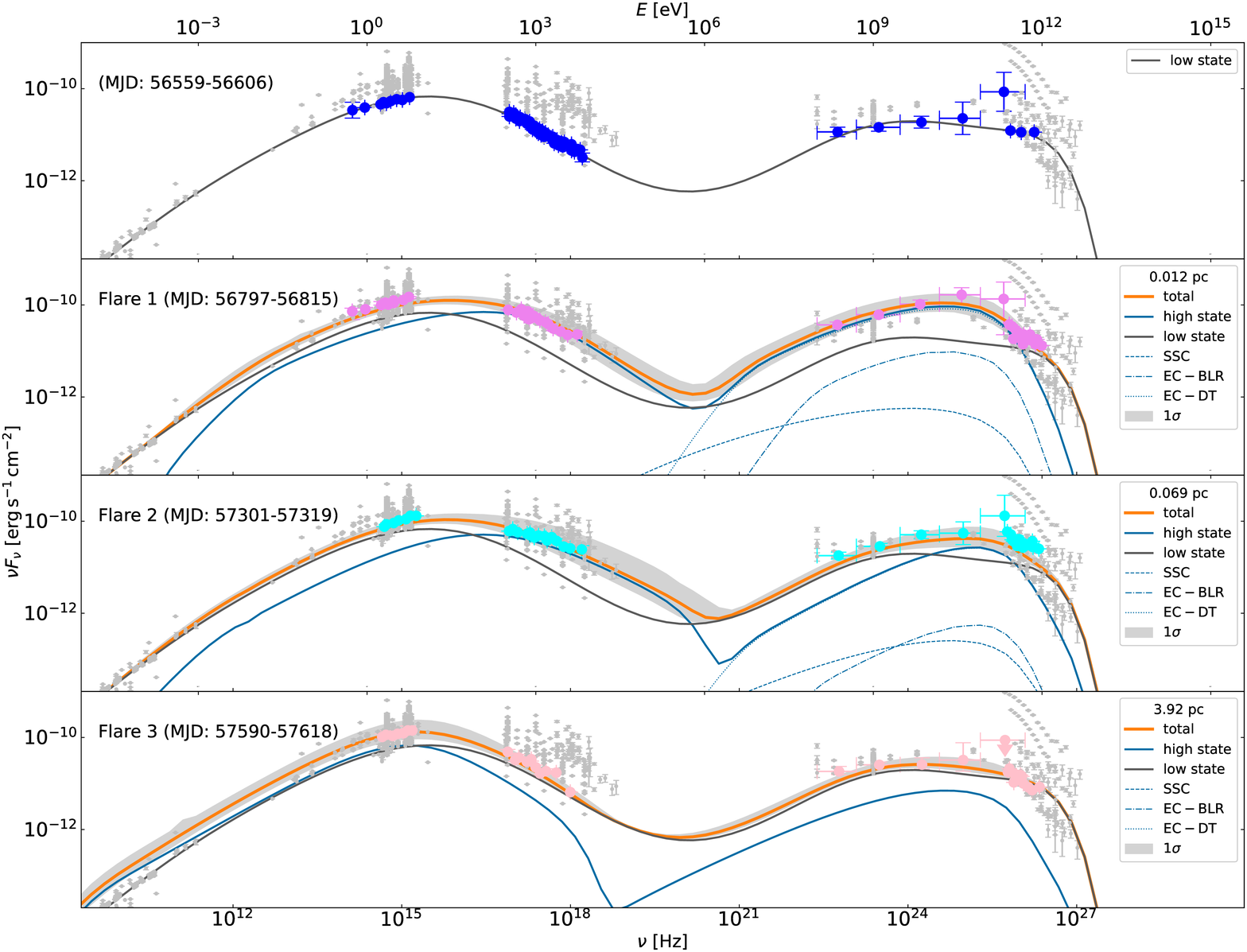}
\caption{Same as Fig.~\ref{fig:3C 279} but for PKS~2155-304.\label{fig:PKS 2155-304}}
\end{figure*}
3C~279 is a very bright and highly variable blazar at all wavelengths. It is classified as an FSRQ at redshift of 0.536. An orphan $\gamma$-ray flare was reported on 20 Dec 2013 \citep{2019ApJ...884..116L}. PKS~2155-304 is a well-known blazar in the southern hemisphere and also has bright and variable emissions, particularly in $\gamma$-ray energies. It is a relatively nearby high synchrotron-peaked (HSP) BL~Lac object at redshift of 0.116. An orphan optical flare lasting a few months was reported for PKS~2155-304 in 2016 \citep{2019hepr.confE..27W}. In addition to the orphan flares, many multi-wavelength correlated flares are observed in both sources. Thus, they are ideal test beds for our model.

\begin{figure*}[htbp]
\centering
\includegraphics[width=2\columnwidth]{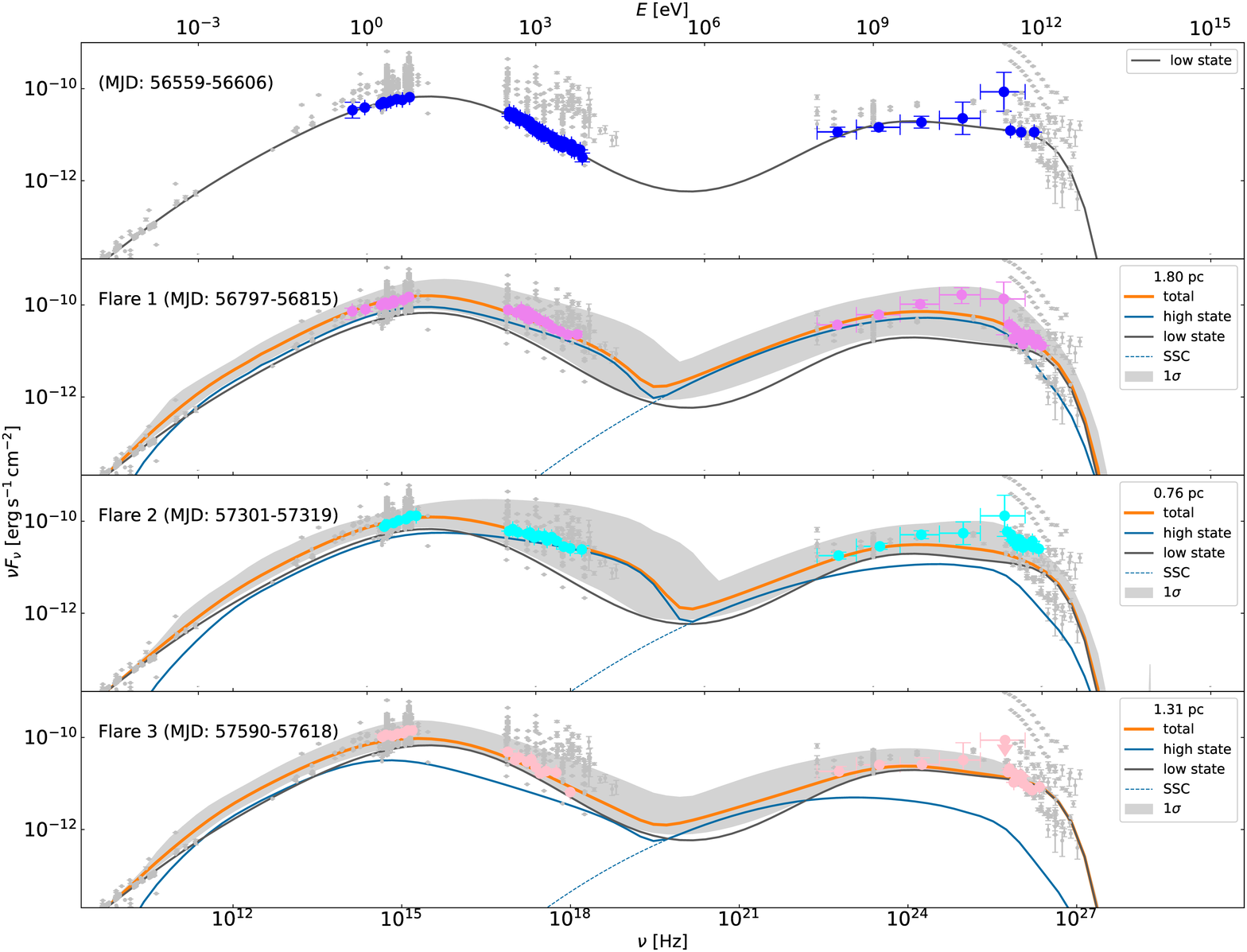}
\caption{Same as Fig.~\ref{fig:PKS 2155-304} for PKS~2155-304 but without external photon field. The common parameters for all flares are: $R_0=7.8^{+1.7}_{-0.9}\times10^{14}\,$cm, $B_0=0.38^{+0.07}_{-0.06}\,$G, $\gamma_{\rm min}=215^{+216}_{-178}$, $\gamma_{\rm max}=1.09^{+0.33}_{-0.61}\times10^6$, $p_1=1.6^{+0.1}_{-0.1}$, $\delta_{\rm D,0}=136^{+16}_{-23}$.  The parameters for each flare read: $L_{\rm e}^{\rm inj}=1.15^{+0.40}_{-0.46}\times10^{45}\,{\rm erg~s^{-1}}$, $p_2=3.7^{+0.3}_{-0.2}$, $\gamma_{\rm break}=1.56^{+0.48}_{-0.43}\times10^{4}$ for Flare 1; $L_{\rm e}^{\rm inj}=3.33^{+1.33}_{-1.01}\times10^{44}\,{\rm erg~s^{-1}}$, $p_2=3.3^{+0.2}_{-0.3}$, $\gamma_{\rm break}=7.49^{+4.63}_{-4.63}\times10^{3}$ for Flare 2; $L_{\rm e}^{\rm inj}=5.77^{+2.81}_{-0.27}\times10^{44}\,{\rm erg~s^{-1}}$, $p_2=3.7^{+0.2}_{-0.1}$, $\gamma_{\rm break}=6.12^{+2.45}_{-6.12}\times10^{3}$ for Flare 3. \label{fig:PKS 2155-304_2}}
\end{figure*}
First, in order to define a low state for each blazar, we search the archival data of each blazar for a period with simultaneous multi-wavelength data of the lowest flux level. We fit the SED of this non-flaring period phenomenologically with a polynomial function and regard it as the background emission component. Then, we choose three flaring states for each blazar that are characterized by different multi-wavelength spectral properties. One of the three flaring states is an orphan optical or $\gamma$-ray flare, and the other two are multi-wavelength flares with different Compton dominance ($q$). It is worth noting that the definition of the orphan flare strongly depends on how the referenced non-flaring SED of the blazar is chosen. For example, Ref.~\citep{2019ApJ...884..116L, 2019hepr.confE..27W} define the orphan flares by comparing the SED of the flaring state with that of the pre-flare state. We here choose the historically lowest-state SED of the blazar as the background emission. Therefore, the reported orphan flares may appear as multi-wavelength flares when compared with the SED of our chosen non-flaring emission.

We search the parameter space, which is composed of eighteen parameters, to find the best-fit values of the model parameters for the three flaring states at once. Note that the multi-wavelength SED of a flare is the superposition of the background component and the flare component, as mentioned in the previous section. The synchrotron radiation and the IC radiation are calculated using the \texttt{naima} Python package \citep{naima}. The best-fit model parameters for each flaring state are obtained with a Markov chain Monte Carlo (MCMC) method (1$\sigma$ error-bars are also obtained with the MCMC method). We use the \texttt{emcee} Python package (version 3.0.2) \citep{2013PASP..125..306F} that is based on the chi-squared statistics. To save computation time and ensure faster convergence, we firstly perform an ``eye-ball fitting'' to the SED and exclude some inappropriate parameter space. We fit the data with 150 parallel walkers for 1000 steps each with a burn-in phase of 300 steps. The parameters and best-fit parameter values including their 1$\sigma$ error-bars can be found in Table~\ref{tab:parameters} and the details of the fitting results of the two blazars will be detailed respectively in the following two subsections. 

\subsection{3C 279}\label{sec:sec4.1}
Fig.~\ref{fig:3C 279} shows the data and best-fit models of four different states of 3C~279. The grey points are historical data which come from the SSDC SED builder\footnote{https://tools.ssdc.asi.it/SED/}. The blue points in the first panel are low-state data collected from February to May of 2010 (period H in Ref.~\citep{2012ApJ...754..114H}). The violet points in the second panel show the SED of the orphan $\gamma$-ray flare (Flare 1) in 2013 as reported by Ref.~\citep{2015ApJ...807...79H}. The cyan and pink points in third and fourth panels are multi-wavelength flaring state data collected on 16 June 2015 (Flare 2) and 1-8 June 2011 (Flare 3), respectively \citep{2019ApJS..245...18F}. The optical flux of the latter two flares is comparable but the $\gamma$-ray flux of Flare 3 is significantly lower than that of Flare 2. The black curve in each panel is the polynomial function characterizing the background emission (low state) component. The solid blue curve shows the flare emission (high state) component. The solid orange curve represents the sum of these two components. The best-fit parameters of 3C~279, which are listed in Table~\ref{tab:parameters}, are within a reasonable range, except for the large Doppler factor $\delta_{\rm D,0}=70.6_{-3.4}^{+4.5}$. Nonetheless, this value is consistent with other studies. For instance, Ref.~\citep{2004AJ....127.3115J} suggested that the Doppler factor of 3C~279 was at least 39 close to the SMBH by analysing the observation results from the Very Long Baseline Array (VLBA). Ref.~\citep{2017AIPC.1792e0015H} found that a very high bulk Lorentz factor ($>50$) at the jet base was required to explain the minute-scale variability of 3C~279 by considering a standard EC model with conical jet geometry. 
Recently, Ref.~\citep{2020MNRAS.492.3829L} found that the bulk Lorentz factors of some moving emission features of 3C~279 should exceed 37 by analyzing VLBA images at 43 GHz. These authors argue that turbulent motions at the relativistic sound speed could boost the Doppler factor up to $\sim70$ when such turbulent velocities are directed toward the line of sight relative to the systemic flow.
As expected, the position of the flaring zone for the orphan $\gamma$-ray flare is the closest to the SMBH with $r=0.2\,$pc, while the ratio between the synchrotron flux and the EC ($\gamma$-ray) flux increases as the distance $r$ increases. A recent paper studied three orphan $\gamma$-ray flares from three FSRQ sources including 3C~279 \citep{2020arXiv201210291P}. Using a two-zone leptonic model, these authors showed that the orphan $\gamma$-ray flare of 3C~279 might have originated from the region close to its SMBH. Even though the flare reported in that paper is not the same as the one studied here, their conclusions about the production site of the orphan $\gamma$-ray flare are consistent with ours.

For FSRQs, the $\gamma$-ray emission mainly arises from the EC process if the flaring zone is relatively close to the SMBH. While the energy density of the magnetic field ($u_{B}(r)\propto B^2(r)$) decreases along the jet as $r^{-2}$, the energy densities of external photons drop more quickly (i.e., $u_{\rm BLR}(r)\propto r^{-3}\delta_{\rm D}^2, u_{\rm DT}(r)\propto r^{-4}\delta_{\rm D}^2$) once the distance is beyond the characteristic radius of BLR or DT ($r>r_{\rm BLR (DT)}$). Therefore, the Compton dominance, considering the BLR and DT components separately, reads $q_{\rm BLR}(r)=u_{\rm BLR}(r)/u_{B}(r)\propto r^{-1}\delta_{\rm D}^2$, and $q_{\rm DT}(r)=u_{\rm DT}(r)/u_{B}(r)\propto r^{-2}\delta_{\rm D}^2$, with $\delta_{\rm D}$ being constant or decreasing with radius. The KN effect would slightly modify the expressions of the Compton ratio but it would not alter the radial dependence of the trend. Such a result suggests that the $\gamma$-ray emission is more intense when the dissipation zone is located closer to the BLR \cite[e.g.][]{2009ApJ...704...38S}, and verifies our speculation that orphan $\gamma$-ray flares tend to appear when dissipation occurs comparatively close to the SMBH. Although no orphan optical flare has been discovered from 3C~279 yet, we may expect to observe orphan optical flares from the source if the flaring zone is located far from the SMBH. 

\subsection{PKS~2155-304}\label{sec:sec4.2}
Fig.~\ref{fig:PKS 2155-304} shows the data and our best-fit models of four different states of PKS~2155-304. The gray points are the sum of historical spectral data. The blue points in the first panel are non-flaring data collected in 2013. The violet and cyan points in the second and third panels show the spectra during a multi-wavelength flare reported in 2014 (Flare 1) and 2015 (Flare 2) respectively. The pink points in the fourth panel give the spectrum measured during an orphan optical flare in 2016 (Flare 3)~\citep{2019hepr.confE..27W}. Note that the very high energy data of PKS~2155-304 are EBL corrected, so we only consider the $\gamma$-ray opacity due to the radiation of the blazar jet.

As a BL~Lac object, PKS~2155-304 is not expected to have strong BLR and DT radiation. Indeed, no emission line is observed in its spectrum, posing an upper limit of $1.1\times 10^{41}~\rm erg~s^{-1}$ on its BLR luminosity. Therefore, we consider two cases for the external radiation field in the modeling of PKS~2155-304. In the first case, we assume the presence of the BLR with the luminosity equal to the measured upper limit, leading to a characteristic BLR radius $r_{\rm BLR}=10^{-3}\,$pc  and a small dust torus in this case. In the second case, we simply do not take into account any external radiation field. The results are shown in Figs.~\ref{fig:PKS 2155-304} and \ref{fig:PKS 2155-304_2}, respectively. All displayed curves are the best-fit models and have the same meaning as those in Fig.~\ref{fig:3C 279}. 
\begin{figure*}[htbp]
\centering
\resizebox{\hsize}{!}{\includegraphics{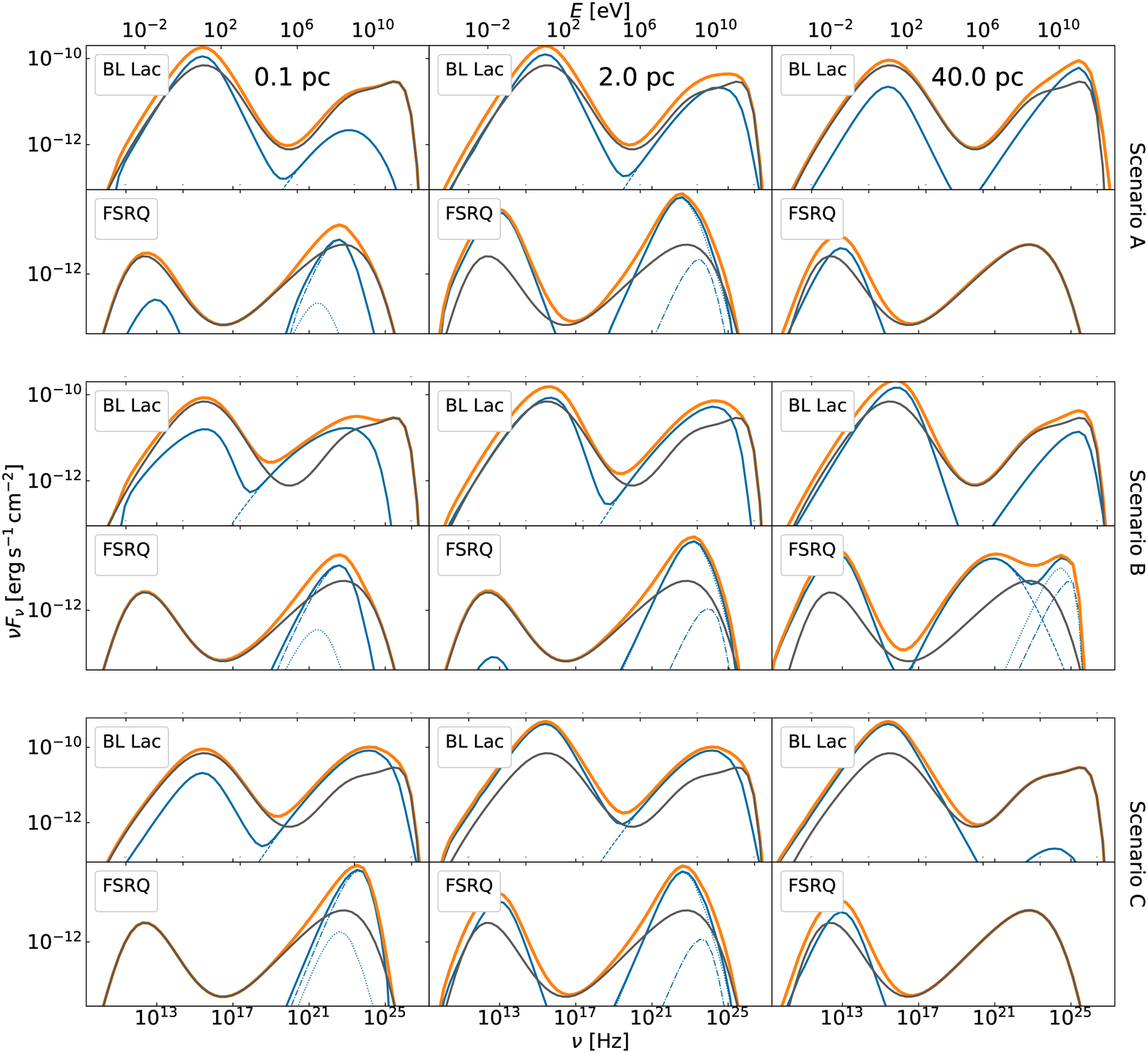}}
\caption{Results for three toy scenarios for orphan and multi-wavelength flares from a fiducial BL~Lac and FSRQ source. In Scenario A (decelerating jet case) and B (accelerating jet case), the magnetic field strength and Doppler factor vary along the jet, while in Scenario C both parameters are considered constant.} The black solid lines describe the low-state SEDs of PKS~2155-304 and 3C~279 that are used as reference for the fiducial BL~Lac and FSRQ, respectively. All lines have the same meaning as in Fig.~\ref{fig:PKS 2155-304}. \label{fig:1}
\end{figure*}

In both cases, the multi-wavelength SEDs in all three flaring states can be satisfactorily reproduced. In the case with external radiation fields, we find a similar trend of the ratio between the synchrotron flux to the IC flux as a function of $r$ as in 3C~279. The orphan optical flare (Flare 3) arises when the flaring zone occurs at a distance (i.e., $r=3.92\,$pc) far beyond the BLR, the characteristic radius of which is $r_{\rm BLR}=10^{-3}\,$pc. The obtained Doppler factor ($\delta_{\rm D,0}=53.6_{-7.0}^{+6.6}$) is a bit larger but still consistent with other studies. For instance, Ref.~\citep{2008MNRAS.384L..19B, 2008MNRAS.386L..28G} suggest that a bulk Lorentz factor above 50 is necessary to explain the minute-scale TeV variability of PKS~2155-304.

However, when no external radiation fields are taken into account, the trend breaks down because the $\gamma$-ray emission in this case is produced by the SSC process that depends on the intensity of the synchrotron radiation and the size of the dissipation zone. The KN effect also plays an important role in the SSC-dominated case. As a result, an extremely large Doppler factor $\delta_{\rm D,0}=136.4^{+15.6}_{-22.5}$ is inferred by the fit. Hence, we do not consider this case as a reasonable solution, at least for the flares of PKS~2155-304. The best-fit results do not correspond well to the observed data, which also brings a very large uncertainty. For a detailed analysis, we refer readers to Appendix~\ref{sec:appendixa}. 

\begin{table*}
\caption{\label{tab:parameters2_B}Indicative parameters for the three illustrative scenarios considered for BL~Lac.}
\begin{ruledtabular}
\begin{tabular}{c|ccc|ccc|ccc}
& \multicolumn{3}{c|}{Scenario A}
& \multicolumn{3}{c|}{Scenario B} 
& \multicolumn{3}{c}{Scenario C} \\  
& $0.1\,$pc
& $2.0\,$pc
& $40.0\,$pc
& $0.1\,$pc
& $2.0\,$pc
& $40.0\,$pc
& $0.1\,$pc
& $2.0\,$pc
& $40.0\,$pc
\\ \hline
$L_{\rm disk}$ ($10^{44}~\rm erg~s^{-1}$) 
& \multicolumn{3}{c|}{}     
& \multicolumn{3}{c|}{}     
& \multicolumn{3}{c}{}     
\\ 
$R_0$($10^{16}~$cm)            & \multicolumn{3}{c|}{$1$}     & \multicolumn{3}{c|}{$1$}     &
\multicolumn{3}{c}{$1$}     
\\ 
{$B_0$\footnote{$B(r)=B_0$ is employed in Scenario C.}}(G)                       
& \multicolumn{3}{c|}{$0.3$}   & \multicolumn{3}{c|}{$2$}   & \multicolumn{3}{c}{$0.05$}   \\ 
{$\delta_{\rm D,0}$\footnote{$\delta_{\rm D}(r)=\delta_{\rm D,0}$ is employed in Scenario C.}} 
& \multicolumn{3}{c|}{$60$}
& \multicolumn{3}{c|}{$5$} 
& \multicolumn{3}{c}{$25$}
\\ 
$p_{\rm 1}$    
& \multicolumn{3}{c|}{$1.5$}  
& \multicolumn{3}{c|}{$1.5$} 
& \multicolumn{3}{c}{$1.5$}  
\\ 
$p_{\rm 2}$                    & \multicolumn{3}{c|}{$4.8$} 
& \multicolumn{3}{c|}{$4.8$}
& \multicolumn{3}{c}{$4.8$}    
\\ 
$\gamma_{\rm min}$          
& \multicolumn{3}{c|}{$10$}
& \multicolumn{3}{c|}{$10$}
& \multicolumn{3}{c}{$10$}
\\ 
$\gamma_{\rm break} (10^3)$   
& $7.0$
& $41.0$
& $348.3$
& $20.0$
& $70.4$
& $247.7$
& \multicolumn{3}{c}{26.6}  
\\ 
$\gamma_{\rm max}$ ($10^6$)   
& $2.0$
& $11.7$
& $99.5$
& $2.0$
& $7.0$
& $24.8$
& \multicolumn{3}{c}{7.6}  
\\ 
$L_{\rm e}^{\rm inj} (10^{43}~\rm erg~s^{-1})$    
& $4.1$    
& $48.1$    
& $267.5$ 
& $0.7$    
& $0.7$    
& $0.35$   
& $43.1$    
& $43.1$    
& $2.2$  
\\
\end{tabular}
\end{ruledtabular}
\end{table*}
\begin{table*}
\caption{\label{tab:parameters2_F}Indicative parameters for the three illustrative scenarios considered for FSRQ.}
\begin{ruledtabular}
\begin{tabular}{c|ccc|ccc|ccc}
& \multicolumn{3}{c|}{Scenario A}
& \multicolumn{3}{c|}{Scenario B} 
& \multicolumn{3}{c}{Scenario C} \\    
& $0.1\,$pc
& $2.0\,$pc
& $40.0\,$pc
& $0.1\,$pc
& $2.0\,$pc
& $40.0\,$pc
& $0.1\,$pc
& $2.0\,$pc
& $40.0\,$pc
\\ \hline
$L_{\rm disk}$ ($10^{44}~\rm erg~s^{-1}$) 
& \multicolumn{3}{c|}{$8$} 
& \multicolumn{3}{c|}{$8$}
& \multicolumn{3}{c}{$8$}
\\ 
$R_0$($10^{16}~$cm)            & \multicolumn{3}{c|}{$5$}     & \multicolumn{3}{c|}{$1$}     & \multicolumn{3}{c}{$5$}
\\ 
{$B_0$}(G)                     &   \multicolumn{3}{c|}{$1.8$}   & \multicolumn{3}{c|}{$0.16$}   & \multicolumn{3}{c}{$0.05$}   \\ 
{$\delta_{\rm D,0}$} 
& \multicolumn{3}{c|}{$60$}
& \multicolumn{3}{c|}{$15$}  
& \multicolumn{3}{c}{$25$}
\\ 
$p_{\rm 1}$    
& \multicolumn{3}{c|}{$1.5$}  
& \multicolumn{3}{c|}{$1.5$} 
& \multicolumn{3}{c}{$1.5$}   
\\ 
$p_{\rm 2}$                    & \multicolumn{3}{c|}{$4.8$} 
& \multicolumn{3}{c|}{$4.8$}
& \multicolumn{3}{c}{$4.8$}  
\\ 
$\gamma_{\rm min}$          
& \multicolumn{3}{c|}{$10$}
& \multicolumn{3}{c|}{$10$}
& \multicolumn{3}{c}{$10$}
\\ 
$\gamma_{\rm break} (10^3)$   
& $0.2$
& $1.2$
& $10.0$
& $1.0$
& $3.5$
& $12.4$ 
& \multicolumn{3}{c}{1.9} 
\\ 
$\gamma_{\rm max}$ ($10^6$)   
& $2.0$
& $11.7$
& $99.5$
& $20.0$
& $11.7$
& $99.5$
& \multicolumn{3}{c}{18.6}  
\\ 
$L_{\rm e}^{\rm inj} (10^{43}~\rm erg~s^{-1})$    
& $0.07$    
& $27.3$    
& $211.4$   
& $0.3$    
& $0.3$    
& $15$     
& $1.4$    
& $42.1$    
& $1.4$ 
\\
\end{tabular}
\end{ruledtabular}
\end{table*}

\section{Influence of the Compton dominance on the types of blazar flares}\label{sec:sec4}
To produce an orphan flare in a certain energy band within our model, the flux of the flaring zone in that energy band should significantly exceed the flux of the background emission, while the flux in any other energy band should remain below the background emission by definition. Therefore, the key to producing an orphan flare, provided a sufficient flux from the flaring zone can be produced, is the comparison between the shape of the flaring zone's SED and that of the background emission's SED. The main feature of the double-humped-shaped SED is the relative amplitude of the two humps, which can be described by the Compton dominance $q$. The ratio between the Compton dominance of the flaring zone $q$ and that of the jet's non-flaring emission, which we hereafter denote as $\chi$, determines the type of the blazar flare. Blazars tend to present an orphan $\gamma$-ray flare if $\chi\gg1$. On the contrary, an orphan optical flare would appear if $\chi\ll 1$. If the Compton dominance of the flaring and quiet state is comparable, a multi-wavelength flare is most likely to be produced within our model. We therefore conduct a more general study of the influence of model parameters on the Compton ratio of the flaring zone and the multi-wavelength properties of a blazar flare in this section.

For this purpose, we consider three generic scenarios. The first one is the same as the model introduced in Section \ref{sec:method}, and is referred to as Scenario A. In addition, we consider an accelerating jet case (Scenario B) in which the Doppler factor increases along the jet and we employ $\delta_{\rm D}(r)=\delta_{\rm D,0}\left(r/{0.1~\rm pc}\right)^{0.16}$, following the observation of M87 jet \citep{2019ApJ...887..147P}. In Scenario C, the magnetic field strength $B$ and the Doppler factor $\delta_D$ are assumed independent of the jet radius. For each scenario, an FSRQ (including the so-called ``masquerading'' BL~Lacs, i.e., \cite{2013MNRAS.431.1914G}) and a true BL~Lac (without external radiation fields) will be studied, with the location of the flaring zone being located at 0.1, 2.0 and 40.0\,pc away from the SMBH respectively. The results are displayed in Fig.~\ref{fig:1} and the parameters can be found in Table~\ref{tab:parameters2_B} for BL~Lac and Table~\ref{tab:parameters2_F} for FSRQ. For this example, we intentionally choose the Compton dominance of the jet's background emission to be unity so that $\chi=q$. The parameters in Table~\ref{tab:parameters2_B} and \ref{tab:parameters2_F} are selected to ensure that all three types of flare (multi-wavelength flare and orphan optical/gamma-ray flare appear as much as possible when the flare region is located in three different positions. For example, a smaller $\delta_{\rm D,0}$ may cause the orphan $\gamma$-ray flare to be unable to appear for FSRQs in all of three scenarios.

In Scenario A, the orphan $\gamma$-ray flares in an FSRQ source are more likely to show up when the location of the flaring zone is comparatively close to the SMBH, while the orphan optical flares are apt to appear far away from the SMBH (see second row from top in Fig.~\ref{fig:1}). For a BL~Lac object, the results in Scenario A show the opposite trend (see first row from top in Fig.~\ref{fig:1}): the higher intensity $\gamma$-ray flare arises when the flaring zone is located far away from the SMBH. The reason is the same as discussed for PKS~2155-304 (see also Appendix~\ref{sec:appendixa}). 

In Scenario B, the situation for BL~Lacs is opposite to those in Scenario A and is more akin to the situation of FSRQs in Scenario A. The orphan gamma-ray flares tend to occur at small distance while the orphan optical flares can be found at large distances from the SMBH. On the other hand, the orphan optical flares are hard to appear in FSRQs which is different from that in Scenario A. The Compton dominance ($q$) is still the key to understanding the difference. 
Contrary to the decelerating jet case, the increasing Doppler factor in scenario B
can slow down the decrease of the Compton dominance along the jet. Therefore, it is necessary that the dissipation occurs at a distance very far from the SMBH to produce an orphan optical flare with the Compton dominance much less than unity. A detailed analysis can be found in Appendix~\ref{sec:appendixb}.

The results in Scenario C are similar to that in Scenario B except that the orphan $\gamma$-ray flare can appear at a medium distance ($r=2$~pc) in Scenario B.
It can also be explained by considering the Compton dominance.
The energy density of the magnetic field is fixed in Scenario C. So, the Compton dominance $q(r)$ decreases with distance as $r^{-3}$ (or $r^{-4}$) for the BLR (DT) if the flaring zone is located beyond the characteristic distances of the BLR (DT) in an FSRQ. For a BL~Lac object, the high-energy emission arises from the SSC process. To focus on the influence of the flaring zone's position (i.e., $r$) on the Compton dominance, let's assume a fixed synchrotron luminosity for the flare. In this case we can obtain the energy density of the synchrotron radiation $u_{\rm syn}\propto r^{-2}$. Thus, the Compton dominance $q(r)$ decreases as $r^{-2}$. The factor of KN effect ($f_{\rm KN}$) remains unchanged given a fixed magnetic field strength and Doppler factor. If the synchrotron luminosity of the flaring zone also decreases with increasing distance, it would lead to a faster decline of the Compton dominance $q(r)$ with respect to $r$. As a result, it becomes more difficult to generate $\gamma$-ray flares at a larger distance.

\section{Discussion}\label{sec:sec5}
\subsection{Duration of blazar flares}
Although we mainly focus on the SED of blazar flares in the present work, the flare duration is another important property. We here briefly discuss the expectations in our model. The duration timescale of a flare cannot be shorter than the light-crossing time of the dissipation zone in the observer's frame, i.e., $t_{\rm lc}\sim(1+z)R(r)/(c\delta_{\rm D})$, which depends on the size of the dissipation zone $R$ or its distance $r$ from the SMBH. In reality, the particle radiative cooling timescale $t_{\rm cool}$, adiabatic timescale $t_{\rm ad}$ or the escape timescale $t_{\rm esc}$ may determine the flare duration if they are longer than $t_{\rm lc}$. Similarly,
the particle acceleration timescale $t_{\rm acc}$ or the injection timescale $t_{\rm inj}$ of accelerated particles into the radiation zone could 
affect
the flare duration.
These timescales depend on the mechanism that triggered the dissipation, which is not specified in our work. Hence, for the following discussion, we focus on the radiative loss and escape timescales.
According to our setup in Section~\ref{sec:method}, all these timescales are shorter for smaller $R$, thus the light-crossing time can be used as a proxy for the flare duration. 

\begin{table*}[htbp]
\caption{\label{tab:timescale}The light-crossing times and observed duration of flaring states for 3C~279 and PKS~2155-304.}
\begin{ruledtabular}
\begin{tabular}{c|cc|cc|cc}
& \multicolumn{2}{c|}{3C~279}      
& \multicolumn{2}{c|}{PKS~2155-304}
& \multicolumn{2}{c}{PKS~2155-304 (no BLR/DT)}\\
& $t_{\rm lc}$
& $\Delta t_{\rm dur}$
& $t_{\rm lc}$
& $\Delta t_{\rm dur}$
& $t_{\rm lc}$
& $\Delta t_{\rm dur}$ \\
\hline
Flare 1
& $0.17^{+0.05}_{-0.04}$ days
& $0.5$ days\footnote{The observed duration of Flare 1 for 3C~279 is reported by Ref.~\citep{2019ApJ...884..116L}. The others are the approximate duration estimated from light curves of the flares.} (G)\footnote{`G' denotes that this duration is estimated from the GeV band, `O' from the optical, `X' from the X-ray, and `T' from the TeV band.}
& $0.03^{+0.02}_{-0.01}$ days
& $\sim20$ days (X/G)
& $0.08^{+0.10}_{-0.03}$ days
& $\sim20$ days (X/G)\\
Flare 2
& $4.55^{+0.96}_{-0.94}$ days
& $\sim2$ days (G)
& $0.13^{+0.10}_{-0.05}$ days
& $\sim20$ days (T)
& $0.03^{+0.03}_{-0.01}$ days
& $\sim20$ days (T)\\
Flare 3
& $30^{+13}_{-9}$ days
& $\sim23$ days (G)
& $22^{+12}_{-7}$ days
& $\sim108$ days (O)
& $0.05^{+0.06}_{-0.03}$ days
& $\sim108$ days (O)\\
\end{tabular}
\end{ruledtabular}
\end{table*}

In the stochastic dissipation model, orphan optical flares may arise if the flaring zone occurs at a distance far from the SMBH, and the range of distances is from a few parsecs to hundreds of parsecs. On the contrary, orphan $\gamma$-ray flares arise for a small range of distances comparatively close to the SMBH for FSRQs in both Scenarios A and B, as well as for BL~Lac objects in Scenario C. As a consequence, we may expect that the duration of orphan optical flares is generally longer than that of orphan $\gamma$-ray flares, regardless of whether the duration is determined by $t_{\rm lc}$ or the other three timescales. We calculate $t_{\rm lc}$ for each flare studied in Section~\ref{sec:sec3} using the best-fit parameters shown in Table~\ref{tab:parameters} and compare it with the observed duration of the flaring state $\Delta t_{\rm dur}$, which is approximated by the time span between the time a flare's flux rises and drops to half of the peak value (i.e., full width at half maximum of the flare's light curve). The results are shown in Table~\ref{tab:timescale} and we can see that $t_{\rm lc}\lesssim \Delta t_{\rm dur}$ is generally satisfying for all flares. Also, the observed flare duration increases with the size of the dissipation zone derived in our model as expected. Observations of other blazar flares are also consistent with the expectation. For example, the duration of all the other reported orphan optical flares are of month-long scales: about 3 months for PKS~0208-512~\citep{2013ApJ...763L..11C} and a few months for PKS~2155-304~\citep{2019hepr.confE..27W}. In contrast, some of the reported orphan $\gamma$-ray flares show intraday duration: 12 hours for 3C~279~\citep{2019ApJ...884..116L}, about 5 hours for PKS~1222+21~\citep{2016MNRAS.463L..26B}. In contrast, Scenario A for BL~Lacs suggests that the duration for orphan $\gamma$-ray flares could be longer than that of orphan optical flares for BL~Lac objects. This is consistent with the reported orphan $\gamma$-ray flare for the blazar 1055+018, the duration of which is above 100 days~\citep{2017ApJ...850...87M}. Note that this object could be classified as a quasar \citep{1995ApJ...443..578M} based on the rest-frame equivalent width ($8~\mathring{\rm A}$) of C III $\lambda1909$ emission line, but in some literature it is referred to as a BL~Lac object because of its relatively weak emission lines \cite[e.g.][]{2014ApJ...789..135W,2015ATel.7114....1J}.

\subsection{Orphan flare rate}
Ref.~\cite{2019ApJ...880...32L} suggested that the true orphan flare rates were $54.5~\%$ and $20~\%$ for optical and $\gamma$-ray flares respectively by analysing a sample with 107 BL~Lac objects, 64 FSRQs, 4 radio galaxies and 3 unclassified sources. In the stochastic dissipation model, the volume of the jet more likely to produce an orphan optical flare is much larger than that to produce an orphan $\gamma-$ray flare, at least for FSRQs or BL~Lacs with weak BLR radiation. The model would predict a much larger intrinsic rate of orphan optical flares than that of orphan $\gamma$-ray flares, if, for example, dissipation takes place in the jet with an equal probability per unit distance or per unit volume. However, since the magnetic field strength and the radiation field intensity are weaker (i.e., lower radiation efficiency of electrons) at larger distance, only very strong dissipation forming at large distance could manifest itself as a distinct flare. This may explain why the observed rate of orphan optical flares is only $\sim 2.5$ times greater than that of orphan $\gamma-$ray flares, and is also consistent with the large electron injection luminosity for flares occurring at large $r$ as shown in Table~\ref{tab:parameters}. In addition, the observed rates of orphan flares might also imply that the dissipation process tends to occur at smaller distance over larger distance in our model. This is not unreasonable because we may generally expect the jet to be more magnetized at a small distance \citep{2019ARA&A..57..467B, 2021MNRAS.502.1145Z}, and hence instabilities or magnetic reconnection may be better developed, while the magnetic energy might have been already (partially) consumed at large distance due to radiation or adiabatic expansion of the jet.

From Section~\ref{sec:sec4} we see that the Scenario A for BL~Lacs suggests an opposite preference of the location for orphan optical flares and orphan $\gamma$-ray flares with respect to FSRQs and the Scenarios B and C for BL~Lacs. Such a difference arises from the different spatial evolution of the jet's parameters. Therefore, if the orphan flare rate can be obtained from a sample composed only of BL~Lacs, it may be possible to distinguish these three different scenarios for BL~Lacs and study the parameter evolution along the jet.

It is worth noting that the rates of orphan flares quoted above have been determined after accounting for sampling and instrumental sensitivity limitations. To determine the fraction of observed true orphan flares, one must estimate the fraction of observed orphan events which are simply due to limited sensitivity of the telescopes used to monitor the sources in various wavelengths. In the analysis of~\citep{2019ApJ...880...32L} the true fraction of orphan $\gamma$-ray flares was estimated by assuming that the true fraction of multi-wavelength (non-orphan) flares is constant as a function of the brightness of the source, and by comparing the expected number of multi-wavelength flares at infinite instrumental sensitivity to that observed.

\subsection{Jet's background emission}
In our model, we consider that dissipation may occur along the entire jet and form numerous emitting blobs. The sum of the emission from those blobs that are not undergoing intense dissipation constitute the background emission of the jet, which may represent the low-state emission of the blazar. The envisaged scenario somewhat resembles the conical jet model which has been proposed by Ref.~\citep{1979ApJ...232...34B} and further developed in many studies \cite[e.g.][]{1980ApJ...235..386M,1981ApJ...243..700K,1985A&A...146..204G,2006MNRAS.367.1083K, 2012MNRAS.423..756P}, in that both models consider the jet as an extended emission region. The difference lies in the particle injection process: in the conical jet model, particles are injected at the jet base and advected to a larger distance. Re-acceleration of the particles along their propagation is needed in order to compensate for the severe radiative cooling and adiabatic energy loss \citep{2015MNRAS.453.4070P, 2019MNRAS.485.1210Z}. In our model instead, relativistic particles are injected locally at both small and large distances instead of being injected at the jet base and transported to larger distances.

\begin{figure}[htbp]
\centering
\resizebox{\hsize}{!}{\includegraphics{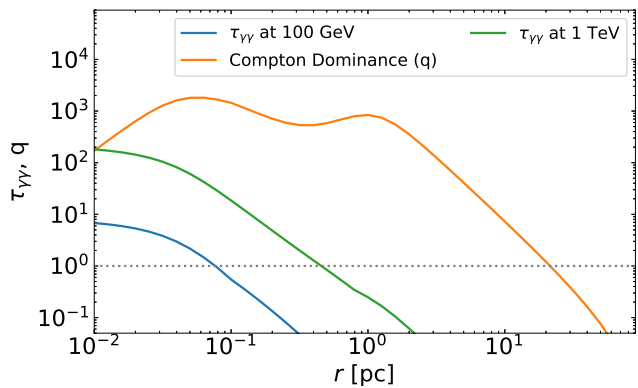}}
\caption{Optical depth, $\tau_{\gamma\gamma}$, and Compton dominance, $q$, as a function of distance from the jet base, $r$. This result is parameter dependent, and here we calculate it with the best-fit parameters for 3C~279. For a much stronger initial magnetic field $q\approx 1$ even inside the BLR. 
\label{fig:VHE}}
\end{figure}
In the previous sections, we ignored the jet's background emission as a target photon field for the IC process of electrons in the flaring zone and for the $\gamma\gamma$ absorption process. To accurately evaluate its contribution, we need to model the distribution of emissivity of the background component along the entire jet, which is beyond the scope of this work. However, it may be safe to ignore their influence in the model. Taking the flares of 3C~279 and PKS~2155-304 for example, we consider the background emission as a target photon field for Compton scattering. For simplicity, we assume that the entire background emission is emitted from the same region of the flaring zone, which significantly overestimates the number density of the background radiation field. We find that the resulting fluxes have little change compared to those shown in Figs.~\ref{fig:3C 279}-\ref{fig:PKS 2155-304_2} for $E_\gamma<1\,$TeV for the same parameters listed in Tables~\ref{tab:parameters}. The most significant change is found in the case of PKS~2155-304 without external photon fields: the flux increases by 30\% around 10\,MeV for flare 3 (where no data is available) and 20\% around 100\,GeV for flare 1 (only one data point from HESS is influenced), and hence does not influence our conclusion. In the presence of an external photon field, which would then dominate the IC process, the influence of the background emission is negligible.

\subsection{Observational tests of the stochastic dissipation model}

\subsubsection{Absorption of the gamma-ray emission}
High-energy $\gamma$-ray photons may not be able to escape from their production site because of the absorption caused by the BLR and the DT radiation via the Breit-Wheeler pair production process. The cross section of the process peaks at 1\,MeV in the center-of-momentum frame. Since the typical photon energy of the BLR radiation is about 10\,eV (i.e. the Lyman-$\alpha$ emission) and the DT radiation is at the infrared band, the absorption is particularly important for photons of energy $\gtrsim 100$\,GeV. Therefore, the location of the emission zone can be determined by searching for the absorption features in the very-high-energy (VHE, energy above 100\,GeV) $\gamma$-ray spectrum. For example, Ref.~\citep{2011ApJ...730L...8A} found that the MAGIC observations of the FSRQ PKS~1222+21 show no spectral cutoff, and concluded that the $\gamma$-ray emission region is located outside the BLR.

The gamma-ray opacity is related to the density of the target radiation field, which is also relevant for the IC emission. We show the opacity $\tau_{\gamma\gamma}$ and the Compton dominance as a function of distance $r$ in Fig.~\ref{fig:VHE} with the best-fit parameters for 3C~279. It can be seen that the Compton dominance approaches unity at jet distance $r\gtrsim 10\,$pc, where the VHE gamma-ray opacity is much smaller than unity. This implies that there should not be an absorption feature at the VHE band in the spectrum of a multi-wavelength flare which has comparable synchrotron flux and IC flux. This is consistent with previous studies \cite[e.g.][]{2014A&A...569A..46A,2021A&A...648A..23H} reporting that there is no such absorption feature during multi-wavelength blazar flares. Furthermore, Ref.~\citep{2019A&A...627A.159H} suggested that the emission zone is confidently beyond the BLR and placed it at $r\gtrsim 1.7\times10^{17}~{\rm cm}$ by fitting the $\gamma$-ray data of 3C~279 observed in June 2015, which corresponds to Flare 2 of 3C~279 in this paper. This is consistent with our fitting results. Although a clear absorption feature in the VHE spectrum of blazar flares has not been reported, our model predicts that such a feature could appear in orphan gamma-ray flares with a high Compton dominance $q\gg 1$. Future observations with next-generation VHE gamma-ray telescopes such as CTA will thus be in a position to test the stochastic dissipation model.

In addition to the VHE $\gamma$-ray photons, even $\gtrsim 10\,$GeV photons may be absorbed by the BLR radiation for high-redshift sources, given favorable conditions (e.g., a very compact radiation zone and intense BLR radiation). Many studies tested this scenario and found no evidence for the expected BLR absorption in the Fermi-LAT spectra \cite[e.g.][]{2012MNRAS.425.2015P,2013MNRAS.435L..24T,2014ApJ...790...45P,2018MNRAS.477.4749C,2019ApJ...877...39M}. On the other hand, there are studies suggesting that the $\gamma$-ray emission must arise within the BLR, because the $\gamma$-ray spectra can not be described by a simple power law for some FSRQ sources \cite[e.g.][]{2010ApJ...717L.118P,2021MNRAS.500.5297A}, although this feature may also be related to the maximum energy of electrons or the KN effect. In any case, observations in dozens of sources in the GeV band is probably also a promising way to study the position of blazars' $\gamma$-ray flares and test the stochastic dissipation model.


\subsubsection{Shift of the radiation center}
\begin{figure}[htbp]
\centering
\resizebox{\hsize}{!}{\includegraphics{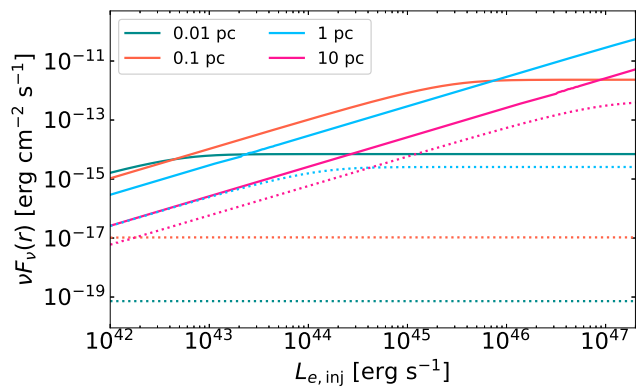}}
\caption{Synchrotron radio flux from the flaring region as a function of electron injection luminosity assuming dissipation at different jet distances. The solid lines represent results at 230~GHz, and the dotted lines represent results at 8~GHz. The parameters are: $z=0.536$, $B_0=0.3\,$G, $R_0=10^{16}\,$cm, $\delta_{\rm D,0}=60$, $p_1=1.5$, $p_2=4.8$, $\gamma_{\rm min}=10$, $\gamma_{\rm break}=1.6\times10^{3}$, $\gamma_{\rm max}=10^{6}$.
\label{fig:7}}
\end{figure}

The stochastic dissipation model suggests that different types of flares can arise when strong energy dissipation occurs at different distances from the base of the jet. Thus, we may expect the shift of the radiation centroid of the jet during the flaring state. Instruments with high spatial resolution may resolve the flaring zone. The Very Long Baseline Interferometry (VLBI) technique in the radio band (i.e. $\sim1-40$~GHz) may reach a sub-milliarcsecond resolution and could provide a decisive test of the model. Ref.~\citep{2017A&A...598L...1K} found that a flare of an AGN induces a change of distance between the apparent jet base and the absolute radio VLBI reference point. In the framework of the stochastic dissipation model, the number of electrons in the strong dissipation zone must be significantly enhanced in order to produce the blazar flare. On the other hand, however, the synchrotron self-absorption (SSA) may be strong and severely attenuate the radio emission of the flaring zone. To be more quantitative, we can write the synchrotron luminosity from a spherical dissipation zone as \citep{2001A&A...367..809K}:
\begin{equation}\label{eq:luminosity}
L(\nu)=4\pi^{2}R^{2}\frac{j(\nu)}{k(\nu)}\left\{1-\frac{2}{\tau^{2}}\left[1-e^{-\tau}\left(\tau-1\right) \right] \right\},
\end{equation}
where $j(\nu)$ is the synchrotron emission coefficient, $k(\nu)$ is the absorption coefficient and $\tau$ is the opacity of the SSA effect. The ratio of $j(\nu)$ and $k(\nu)$ is independent of the electron injection luminosity. So the synchrotron luminosity is proportional to:
\begin{equation}\label{eq:luminosity2}
L_{\rm SSA}(\nu)\propto{R^2B^{-\frac{1}{2}}\nu^{\frac{5}{2}}}.
\end{equation}
in the case of $\tau\gg 1$. Therefore, the radio emission may not be sensitive to the enhanced electron injection luminosity during strong flares for a given $R$ and $B$. This can be seen in Fig.~\ref{fig:7}, which shows the relation between the electron injection luminosity and the radio flux of the flaring region at different radii. We see that at a comparatively low frequency such as 8\,GHz, a huge electron injection luminosity does not significantly enhance the radio flux at small radii (e.g., at 0.01\,pc), but may be revealed when the intense dissipation occurs at a large distance with $r>10\,$pc. Therefore, we may expect an orphan optical flare accompanied by the brightening or emergence of a radio knot at a large distance of the jet. Of course, this also depends on the ratio of the radio flux during the flare to that in the low state. If the low-state radio emission is already quite strong, the shift of the radio center during the flare may not be easy to confirm. 

Observations at a higher frequency can alleviate the SSA effect. At 230\,GHz, as shown in Fig.~\ref{fig:7}, the SSA effect is already unimportant at $r>0.1\,$pc. The Event Horizon Telescope (EHT) can work at this frequency and resolve the innermost jet of 3C 279 at 230 GHz with an angular resolution of $\sim20~{\rm\mu as}$ \citep{2020A&A...640A..69K}, which corresponds to a physical length of about 3.7 pc for a viewing angle of $2^{\circ}$ \citep{2017ApJ...846...98J}. With this resolution, it may resolve the dissipation zones of orphan optical flares and even some multi-wavelength flares in 3C 279. A closer blazar would be a better target for such a study. Note that, even at small radii with $r=0.01\,$pc where orphan $\gamma$-ray flares more likely take place, the 230\,GHz radio flux can be increased by about a factor of 3 during the flare. Although the EHT cannot spatially resolve the flaring zone at such a small distance, it is possible to observe a moderate enhancement of the radio flux in the innermost core during an orphan $\gamma$-ray flare. Again, this depends on the radio flux ratio of a flaring state to the low state.

The space mission $Gaia$ provides highly accurate optical centroid positions for AGN with (sub-)milliarcsecond accuracy. Based on Gaia, a number of publications found that there are significant radio-optical offsets for AGN by analyzing VLBI positions and $Gaia$ photocenter \cite[e.g.][]{2017MNRAS.467L..71P,2017MNRAS.471.3775P,2019MNRAS.482.3023P,2019ApJ...871..143P,2017A&A...598L...1K,2020MNRAS.493L..54K}. We can also try to measure the shift of the $Gaia$ optical photocenter during the flare state to test the stochastic dissipation model. For example, let us assume that the optical photocenter of 3C 279 in the low state is located 12 pc away from the SMBH. In Flare 1, which occurs at 0.2 pc away from the SMBH based on our fitting, the optical flux is about ten times higher than that in the low state (see Fig.~\ref{fig:3C 279}). This translates to a shift of the optical photocenter by $\sim 0.2$ mas (given the source redshift and the viewing angle ($2.1^{\circ}$ from Ref.~\citep{2012ApJ...754..114H}).
Such a shift exceeds the pointing accuracy of $Gaia$ for a bright source and hence is measurable by $Gaia$ \citep{2021MNRAS.505.4616B}.


\section{Conclusions}\label{sec:sec6}
In this paper, we have succeeded in interpreting the spectral variety of blazar flares in a unified physical picture. In the considered framework, dissipation events may take place and accelerate particles along the jet at random distances from the SMBH where the electromagnetic environments can be quite different. As a result, different spectral shapes of the emergent radiation can arise from these dissipation zones at different distances. Our model, which is coined the ``stochastic dissipation model'', and the main conclusions of the paper are summarized below.

(i) In our stochastic dissipation model, there are at least two emission components during a blazar flare. One component is the jet's background emission, which can be thought of as a superposition of radiation from numerous but comparatively weak dissipation zones along the jet. The other component originates from a flaring zone (with stronger dissipation) that is responsible for the flare.

(ii) We assume that the flaring zone may randomly appear at different positions along the jet. The physical quantities describing flares from the same blazar, such as the radius of the flaring zone, the magnetic field strength, and Doppler factor, are not independent, but are intrinsically related to each other through the distance of the flaring zone to the SMBH.

(iii) We have applied our model to explain the SEDs of three flaring states of 3C~279 and PKS~2155-304. The SEDs of different flaring states can be explained by our model with six common parameters and four separate parameters for both PKS~2155-304 and 3C~279. Our model for PKS~2155-304, which is categorized as a BL~Lac object, strongly favors the presence of a weak BLR radiation of luminosity $\sim 10^{41}\rm erg~s^{-1}$ which is consistent with the reported upper limit to the BLR luminosity of this source. 

(iv) The ratio $\chi$ between the Compton dominance of the flaring zone and that of the jet's background emission determines the spectral feature of the blazar flare. If the ratio is much larger than unity, the blazar tends to present an orphan $\gamma$-ray flare; on the contrary, if the ratio is much smaller than unity, an orphan optical flare is more like to occur. 

(v) For FSRQs including ``masquerading'' BL~Lacs, the Compton dominance ratio $\chi$ would be much larger than unity (corresponding to orphan $\gamma$-ray flare) when the dissipation occurs comparatively close to the SMBH (e.g., $r\lesssim 1\,$pc), while the ratio would be much smaller than unity (corresponding to orphan optical flare) when the dissipation occurs far away from the SMBH (e.g., $r\gtrsim 10\,$pc). 

(vi) For (true) BL~Lacs, the situation is similar to that of FSRQs, if the model parameters such as the magnetic field and the Doppler factor do not vary with the distance of the dissipation zone from the SMBH ($r$). On the other hand, if these parameters decrease with $r$, the situation becomes more complex due to the KN effect. A dedicated study is needed to elucidate the influence of the KN effect in the latter case.

(vii) The flare duration and the orphan flare rates expected in the model are consistent with orphan flare observations made to date.

In this work we only consider the radiation of electrons in the jet. In principle, protons can also be accelerated in the dissipation zone and radiate neutrinos via the hadronic interactions with the radiation field in blazars. Indeed, an orphan neutrino flare from TXS 0506+056 was reported by the IceCube Neutrino Observatory \cite{2018Sci...361..147I}. In the work of Ref.~\citep{2020arXiv201103681X} it was shown that the neutrino flare may have been produced by a dissipation event occurring at the jet base where the external radiation field is dominated by the X-ray corona of the SMBH. This interpretation is consistent with our model.

\acknowledgements
We thank the anonymous referee for the enlightening suggestions and Alicja Wierzcholska for her help with the data of PKS~2155-304. The work is supported by the NSFC grants 11625312, U2031105 and 11851304, and the National Key R$\&$D program of China under the grant 2018YFA0404203. MP acknowledges support from the MERAC Foundation. Part of this work is based on archival data, software or online services provided by the Space Science Data Center - ASI. The calculation of radiation is based on the \texttt{naima} Python package, and the data fitting is based on the \texttt{emcee} Python package.

\appendix
\section{The influence of KN effect to the Compton dominance}\label{sec:appendixa}
In this appendix, we will give an analytical discussion about why the model without external radiation field for fitting the SEDs of PKS~2155-304 needs such a large Doppler factor. 

For the case without external photon field, the $\gamma$-ray emission arises from the SSC process. We discuss an indicative case, in which the synchrotron luminosity is fixed along the jet, and the energy density of synchrotron photons decreases along the jet $u_{\rm syn}(r)\propto R^{-2}(r)\delta_{\rm D}^{-2}(r)\propto r^{-2}\delta_{\rm D}^{-2}(r)$. (The synchrotron luminosity may also have other forms of evolution along the jet. There is no observational signature to distinguish.) We take the KN effect into account to calculate the Compton dominance. We assume that the energy distribution of soft photons is a power law $u_{\rm ph}(\nu)\sim\nu^{-\alpha}$, and $\nu_{\rm min}$ and $\nu_{\rm max}$ are the minimum and cut off frequency, respectively. The peak Compton dominance is approximately the ratio of the luminosity of the SSC peak to that of the synchrotron peak $q^{\rm peak}(r)\approx f^{\rm peak}_{\rm KN}u_{\rm syn}(r)/u_{B}(r) = f^{\rm peak}_{\rm KN}\delta_{\rm D}^{-2}(r)$, where the factor $f^{\rm peak}_{\rm KN}$ is defined as
\begin{equation}\label{eq:7}
f^{\rm peak}_{\rm KN}(r)\propto
\begin{cases}
(1+\hat{b}_{\rm KN})^{-3/2}& \text{if $\alpha<-0.5$}\\
\tilde{b}_{\rm KN}^{\alpha-1}& \text{if $-0.5<\alpha<1$ ,}\\
(1+\tilde{b}_{\rm KN})^{-3/2}& \text{if $\alpha>1$}
\end{cases}
\end{equation}
where $\hat{b}_{\rm KN}=4\gamma_{\rm break}h\nu_{\rm max}/m_e c^2$ and $\tilde{b}_{\rm KN}=4\gamma_{\rm break}h\nu_{\rm min}/m_e c^2$ \citep{2005MNRAS.363..954M}. 

We now apply Eq.~\ref{eq:7} to the case of PKS~2155-304. It can be seen in the second and the third panels of Fig.~\ref{fig:PKS 2155-304_2} that these two synchrotron components are similar, but the data show that the flux of the IC component in the second panel is more than twice as high than that in the third panel. Due to the presence of background radiation, the flux around the IC peak produced from the flaring zone in the second panel is more than five times higher than that in the third panel. To find out how can the peak Compton dominance differ so much, we use the synchrotron peak frequency as the characteristic frequency of soft photons, and the KN suppression factor, $b^{\rm peak}_{\rm KN}(r)=4\gamma_{\rm break}h\nu_{\rm syn}^{\rm peak,obs}/m_{\rm e}c^2$ which can be approximated by
\begin{equation}\label{eq:8}
\begin{aligned}
b^{\rm peak}_{\rm KN}(r)&\approx 1.7\times10^{-23}(\nu_{\rm syn}^{\rm peak,obs})^{3/2}\delta_{\rm D}^{-3/2}(r)B^{-1/2}(r)\\
&\propto\delta_{\rm D}^{-3/2}(r)r^{1/2},
\end{aligned}
\end{equation}
where $\nu_{\rm syn}^{\rm peak,obs}\approx3.7\times10^6 \gamma^2_{\rm break}B(r)\delta_{\rm D}(r)/(1+z)$ \citep{1998ApJ...509..608T} is the synchrotron peak frequency in the observer frame. As mentioned above, we ignore the background radiation of the jet as the target photon field of the IC process, so the synchrotron photons from the flaring zone are the only considered soft photons. As can be seen in Fig.~\ref{fig:PKS 2155-304_2}, the spectrum of the synchrotron component can be well described as a broken power law ($\nu F_{\nu}\propto \nu^{p_\gamma}$). The slopes are $p_{\gamma, 1}^{\rm Flare~1}=0.4$, $p_{\gamma,2}^{\rm Flare~1}=-0.3$, $p_{\gamma,1}^{\rm Flare~2}=0.3$ and $p_{\gamma,2}^{\rm Flare~2}=-0.2$, respectively. We are focusing on the photons near the peak ($\nu_{\rm peak}^{\rm Flare~1}=3.4\times10^{15}~{\rm Hz}$, $\nu_{\rm peak}^{\rm Flare~2}=5.7\times10^{15}~{\rm Hz}$), so we treat the spectrum around the peak as a single power law with slope $p_{\gamma}^{\rm Flare~1,2}\approx 0$ for simplicity. Then we can calculate that the index $\alpha$ is approximately equal to 1. We substitute Eq.~\ref{eq:8} into Eq.~\ref{eq:7} and thus we may write
\begin{equation}\label{eq:9}
q^{\rm peak}(r)\propto\delta_{\rm D}^{-1/2-3\alpha/2}(r)r^{\alpha/2-1/2}\propto\delta_{\rm D}^{-2}.
\end{equation}

So the five times IC peak flux difference between the second and the third panels requires a significant difference between $\delta_{\rm D}^{\rm Flare~1}$ and $\delta_{\rm D}^{\rm Flare~2}$, which favors a larger Doppler factor at 0.1 pc. And in this case, the higher intensity $\gamma$-ray flare will arise when the location of the flaring zone is far away from the SMBH. This is consistent with the fitting result derived by the MCMC method.

For comparison, we may also compare to the Compton dominance in the EC-dominated case, i.e., 
\begin{equation}\label{eq:10}
q_{\rm BLR, KN}(r)\propto(1+\hat{b}_{\rm KN, BLR})^{-3/2}\delta_{\rm D}^{2}(r)r^{-1}, 
\end{equation} 
for EC/BLR-dominated case and where $\hat{b}_{\rm KN, BLR}\approx 1.4\times10^{-5} {\gamma_{\rm break}}\delta_{\rm D}(r)$.
\begin{equation}\label{eq:11}
q_{\rm DT, KN}(r)\propto(1+\hat{b}_{\rm KN, DT})^{-3/2}\delta_{\rm D}^{2}(r)r^{-2}, 
\end{equation}
for EC/DT-dominated case and where $\hat{b}_{\rm KN, DT}\approx 9.6\times10^{-7} {\gamma_{\rm break}}\delta_{\rm D}(r)$. Substituting the best-fit parameters of 3C 279 into Eq.~\ref{eq:10} and Eq.~\ref{eq:11}, we find that the KN effect would only slightly modify the Compton dominance.

\section{Influence of the form of $\delta_D(r)$} \label{sec:appendixb}
According to the discussion in Appendix~\ref{sec:appendixa}, we see that the influence of the Doppler factor on the Compton dominance is important. Although we employ Eq.~(\ref{eq:4}) to describe the spatial evolution of the Doppler factor, we note that this form is simply an assumption. While our model is flexible so that other forms of $\delta_D(r)$ can be easily substituted into, the realistic form of $\delta_D(r)$ depends on detailed modelling/simulations of jet's dynamics from small to large scales. So we here briefly discuss the influence of the form of $\delta_D(r)$ on the production of orphan flares, which basically boils down to the impact of $\delta_D(r)$ on the Compton dominance $q$. 

For an EC-dominated case, the Compton dominance of a $\gamma$-ray flare will depend quadratically on $\delta_D$, as Eqs.~(\ref{eq:10}) and (\ref{eq:11}) show. Thus, the choice of a $\delta_D(r)$ decreasing faster with $r$ than the adopted one (see Eq.~(\ref{eq:4})) would result in a faster decrease of the Compton ratio than the one found in our results (see e.g., Fig.~\ref{fig:1}). In this case, the region of the jet being suitable for the production of orphan $\gamma$-ray flares would shrink, while that of orphan optical flares would enlarge. On the other hand, if the SSC process dominates the $\gamma$-ray emission, the situation becomes quite complicated. In this case, the dependence of the Compton ratio $q$ on the Doppler factor is less straightforward due to the impact of KN effects and the dependence on the spectral shape of the synchrotron photons (see Eqs.~\ref{eq:7} and \ref{eq:9}). More specifically, depending on the spectral index of the target photon field, the relation between $q(r)$ and $\delta_D(r)$ could be either positive or negative. As a consequence, the influence of the form of $\delta_D(r)$ depends on specific circumstance of the source. For the case of PKS~2155-304, as shown in Eq.~\ref{eq:9}, a faster decreasing $\delta_D$ with $r$ would lead to the opposite trend expected for the EC-dominated case: the region suitable for the production of orphan $\gamma$-ray flares would enlarge at the expense of the orphan optical flare region.

The Doppler factor can also increase as $r$ increases in the case of an accelerating jet. As shown in Fig.~\ref{fig:1}, the orphan $\gamma$-ray flare tends to appear at a smaller distance in both the SSC-dominated case and the EC-dominated case. Another point worth emphasizing is that, if the EC process dominates the $\gamma$-ray emission, the region viable for the production of orphan $\gamma$-ray flares is larger than that in the decelerating jet case. This is consistent with Eq.~\ref{eq:10} and \ref{eq:11}, the Compton dominance tends to decrease more slowly or even increase if the Doppler factor increases along the jet.

\bibliographystyle{apsrev}
\bibliography{references.bib}
\end{document}